# Evidence for Ubiquitous Strong Electron-Phonon Coupling in High Temperature Superconductors.

A. Lanzara*†, P. V. Bogdanov*, X. J. Zhou*, S. A. Kellar*, D. L. Feng*, E. D. Lu†, T. Yoshida‡, H. Eisaki*, A. Fujimori‡, K. Kishio§, J. -I. Shimoyama§, T. Noda||, S. Uchida||, Z. Hussain†, and Z.-X. Shen*.

*Department of Physics, Applied Physics and Stanford Synchrotron Radiation Laboratory, Stanford University, Stanford, CA 94305.

†Advanced Light Source, Lawrence Berkeley National Lab., Berkeley, CA 94720.

‡Department of Physics, University of Tokyo, Bunkyo-ku, Tokyo 113, Japan.

§Department of Applied Chemistry, University of Tokyo, Tokyo 113-8656, Japan.

||Department of Superconductivity, University of Tokyo, Yayoi2-11-16, Bunky-ku, Tokyo 133, Japan.

**Coupling between electrons and phonons (lattice vibrations) drives the formation of the electron pairs responsible for conventional superconductivity[1]. The lack of direct evidence for electron-phonon coupling in the electron dynamics of the high transition temperature superconductors has driven an intensive search for an alternative mechanism. A coupling of an electron with a phonon would result in an abrupt change of its velocity and scattering rate near the phonon energy. Here we use angle resolved photoemission spectroscopy to probe electron dynamics - velocity and scattering rate- for three different families of copper oxide superconductors. We see in all of these materials an abrupt change of electron velocity at 50-80meV, which we cannot explain by any known process other than to invoke coupling with the phonons associated with the movement of the oxygen atoms. This suggests that electron-phonon coupling strongly influences the electron dynamics in the high-temperature superconductors, and must therefore be included in any microscopic theory of superconductivity.**

We investigated the electronic quasiparticle dispersions in three different families of hole-doped cuprates, $Bi_2Sr_2CaCu_2O_8$ (Bi2212) and Pb doped Pb-Bi2212, Pb-doped $Bi_2Sr_2CuO_6$ (Pb-Bi2201) and $La_{2-x}Sr_xCuO_4$ (LSCO). Except for the Bi2201 (overdoped, $T_c$=7K) data and that in Fig. 3b, recorded at the beam-line 5.4 of the Stanford Synchrotron Radiation Laboratory (SSRL), all the data were recorded at the Advanced Light Source (ALS), as detailed elsewhere[2].

The top panels of figure 1 report the momentum distribution curve (MDC) derived dispersions along the (0, 0)-($\pi$, $\pi$) direction for LSCO (panel a) and Bi2212 (panel b) superconducting state and for Pb-Bi2201 normal state (panel c) vs the rescaled momentum, k', defined by normalizing to one the momentum k relative to the Fermi momentum $k_F$, (k-$k_F$), at the binding energy E=170meV. A "kink" in the dispersion around 50-80meV, highlighted by thick arrows in the figure, is the many-body effect of interest and has been reported before[2-5]. The direct comparison suggests a similar phenomenon in different systems (although details differ) and at different doping with the effect gets stronger in the underdoped region. These results put a strong constraint on the nature of the effect. The similar energy scale of the excitation (50-80 meV) in systems with very different gap energy, ranging from (10-20 meV) for LSCO and Bi2201, to (30-50meV) for Bi2212, rules out the superconducting gap as the origin. The data also rule out the proposed explanation[6] in terms of coupling with the magnetic mode at 41meV[7], because of its temperature dependence and its ubiquity. The phenomenon is observed in LSCO where the magnetic mode does not exist. It is also observed well above the transition temperature in all cases but the magnetic mode sets in at $T_c$ for optimal and overdoped samples[7] (e.g. the 30K data from overdoped Pb-Bi2201 are measured at six times $T_c$). These considerations, based on direct experimental evidences, leave phonons as the only possible candidate.

The phonon interpretation receives very strong support from a direct comparison between photoemission results and neutron scattering data on LSCO. As shown by the red arrow and the shaded area in Fig.1a, the energy of the zone boundary in-plane oxygen stretching longitudinal optical (LO) phonon, identified by neutron as being strongly coupled to charge[8, 9], coincides with the "kink" energy in our data baring minor correction with the small superconducting gap of 10meV. We identify this mode as the highest energy phonon that strongly contributes to the ARPES data. However, other phonon contribution should be considered as well, for example the apical mode[10] related with the in plane oxygen-stretching mode. Additional support to the phonon scenario comes also from a temperature dependence analysis. In the lower panel of Fig. 1 we report the evolution of the kink above and below the transition temperature for LSCO (Fig. 1d) and Bi2212 (Fig. 1e). In both cases the kink persists clearly above the transition temperature as expected in the case of el-ph interaction, however a thermal broadening is present.

Fig. 2a-c compares energy distribution curve (EDC) data of Bi2212 along (0, 0)-(π, π), directions (panel a-c) at different dopings, with that of Be (0001) surface state[11, 12] (Fig. 2d) -in which the el-ph interaction is known to be strong - and with simulated spectra (Fig. 2e). A strong similarity among the raw spectra can be observed. In the case of Be, the 65meV phonon sets a scale below which a sharp quasiparticle is possible. This leads to a "dip" or "break" in the spectra near the phonon energy, indicated by the vertical dashed line. This feature is well reproduced in the simple simulation, where a quasiparticle is isotropically coupled to a collective mode near 70meV[6, 13]. The Bi2212 data show also a clear resemblance, with a (π, 0) LO-phonon near 55 meV[9], but pushed up by the superconducting gap. All the data suggest that there is an energy scale in the problem close to expected phonon frequency. Therefore, the striking similarity between Bi2212 data and that of a known phonon case strongly supports the phonon explanation.

In Fig. 3a we report the quasiparticle width of Bi2212 for different doping, using a typical procedure[2]. As shown before[2, 3] the width decreases more rapidly below the relevant energy, which corresponds to the kink position in the dispersion. This again is consistent with the case of an el-ph coupled system, complementing dispersion data in Fig. 1. The drop in the scattering rate near the relevant energy is also seen in optics experiments[14]. Fig. 3b reports the "dip" energy of ARPES spectra near (π, 0) as a function of doping. The data are consistent with the idea that the 55 meV LO phonon energy sets the lower bound[9]. A realistic model, which considers the change of the (π, 0) band position with doping, bilayer splitting, which severally distorts (π, 0) spectra[15, 16], el-ph interaction and energy gap, will be needed to understand the specifics of the data, including the small decrease of the dip energy with doping.

In a simple el-ph model the ratio of the velocity change, between the bare and the dressed electrons velocity is ($\lambda$+1), where $\lambda$ measures the el-ph coupling strength. We can extract a similar quantity from the ratio between the high-energy velocity above the phonon energy (which approximates the velocity without phonon) and the dressed velocity below the phonon energy. The two velocities are determined by fitting the experimental dispersions with two straight lines. The coupling strength determined from the experimental data, $\lambda'$, is proportional to the el-ph coupling constant $\lambda$ but an overestimate for two reasons. The first is that the high-energy dispersion overestimates the bare velocity. The second is that the electron-electron scattering also influences the dispersion away from the phonon energy, especially in the high-energy part that we use as the bare velocity. In Fig.1f we report the doping dependence of $\lambda'$ along the $(\pi, \pi)$ direction. An estimate of $\lambda$ can alternatively be obtained by fitting the experimental data with a model (e.g, Debye model), or assuming the bare velocity as given from band calculations. In both cases the obtained values are smaller than $\lambda'$, but the decreasing trend with doping is identical.

In figure 4a, we show the velocity ratio for different materials, as a function of the angle $\phi$ along the entire FS (panel b). Data obtained from different experimental runs are included in the figure (full and empty symbols). In panel b) we report the rescaled MDC derived dispersions for Pb-Bi2212 taken with 55 eV photon along the two high symmetry directions, $\Gamma Y$ $((0, 0)$-$(\pi, \pi))$ and $\Gamma M$ $((0, 0)$-$(\pi, 0))$. The total change in the ratio with angle is slightly less than a factor of two. In particular, the data from both Pb-doped Bi2212 and Pb-doped Bi2201, where the superstructure effect near $(\pi, 0)$ is minimized, gives confidence to the conclusion. The conversion from velocity ratio to the coupling constant $\lambda$ away from the $(\pi, \pi)$ direction requires the correction due to band structure effect including bilayer-splitting that increases away from $(\pi,\pi)$ direction[15, 16], but the conclusion that the coupling is not very anisotropic remains true. No direct relation with the superconducting transition is observed, as can be seen in the same figure where data above (100K) and below (20K) the transition temperature are reported in the case of Pb-doped Bi2212. Both the angular and the temperature dependence are consistent with the phonon interpretation, as phonons are more isotropic and will not disappear above $T_c$.

Although these experimental findings points toward phonons as the only possible scenario, some literature needs to be addressed. The kink in the dispersion along the (π, π) direction and the dip in the EDCs near (π, 0) has been attributed[6] to the magnetic mode of 41 meV[7]. Comparing the doping dependence of the energy difference between the dip and the superconducting gap in photoemission data, with that of the magnetic mode in neutron data, a definitive proof for the magnetic mode as the origin of the "peak-dip-hump" structure has been claimed[17]. On the other hand, it has been argued for the absence of an energy scale in the problem[4] and the kink in the dispersion was overlooked[4]. The curving in the normal state dispersion has been attributed to the marginal Fermi liquid (MFL) self-energy[18], and the presence of a sharp kink below $T_c$ to coupling to the (π, π) magnetic mode[19].

Apart from its incompatibility with our data, the magnetic mode scenario has serious weaknesses. First, the mode contains only few percent of the total spin fluctuation spectral weight[20], unlikely an explanation for the very large change seen in EDC spectra above and below $T_c$. Second, the electronic model is difficult to implement self-consistently as the mode will be altered by the interaction with electrons[6]. Third, the magnetic mode calculation claims to reproduce the experimental data that yield a huge change of the velocity ratio near (π, 0) region[6]. The inset of Fig. 4a compares the velocity ratio found in the present study with data by previous report[3]. The comparison shows good agreement for ϕ≤30 degree, but a large discrepancy is observed for ϕ≥30 degree. We attribute this difference to confusion caused by the superstructure, which complicates the data near (π, 0) region in pure Bi2212 sample used by previous study[3]. We have chosen superstructure free Pb-Bi2212 for this investigation.

Our data and the above discussion rule out the magnetic mode as a possible scenario, and this makes also the MFL plus magnetic mode interpretation an unnatural explanation for the data. Furthermore, some of the normal state dispersion, for example the 7% underdoped LSCO (fig. 1a) measured at $T_c$ where the kink is sharper, cannot be fit within the MFL approach, even with an arbitrary choice of parameters. If the 7% LSCO data can only be explained by coupling to phonons, it is likely that a similar effect, observed at higher doping and temperatures, requires the same cause. Finally, this attempt to attribute the "kink" in the dispersion, occurring at the same energy above and below $T_c$ to two different origins is not consistent with the data in Fig. 1d and Fig. 1e where one sees that the effect washes out gradually with temperature.

The phonon model can explain all the aspects of the data attributed to the magnetic mode in a more natural way (N. Nagaosa *et al*., personal communication). This approach has the added advantage of not suffering from the tiny spectral weight and the self-consistency issue present in the magnetic mode calculation. The phonon picture naturally explains the gradual temperature evolution observed in the MDC derived dispersions. Although el-ph interaction provides the only interpretation consistent with overall body of data, a few caveats needs to be pointed out. Standard theory of el-ph coupling would produce a drop in resistivity at phonon frequency $\Omega$, or a saturation above a temperature T ~ 0.3-0.5$\Omega$. The drop at $\Omega$ is observed and discussed in connection with Fig. 3, although it was given an alternative interpretation[22]. The temperature dependence of the resistivity shows a complex doping dependence, with a saturation seen in underdoped regime but not near optimal and overdoped regime[22]. An understanding of this complex behavior requires going beyond the simple Fermi liquid theory, but is not incompatible with the presence of el-ph coupling[24]. The lack of resistivity saturation is also seen in doped C60 compound where the el-ph interaction is very strong[24]. An improved theory needs to consider the strong electron-electron interaction, the pseudogap[25], and vertex correction that suppress the large momentum transfer scatterings[26], making phonons hard to detect by resistivity experiments. Finally, we stress that el-ph interaction in strongly correlated materials is a largely unexploited topic, and we hope that our finding can stimulate more theoretical work.

We aknowledge N. Nagaosa, D. J. Scalapino, R. Laughlin, D.-H. Lee, S. Kivelson, D. Bonn, K. A. Muller, P. Allen, N. P. Armitage, A. Damascelli and F. Ronning for useful discussion. The work at ALS was supported by DOE's Office of Basic Energy Science, Division of Materials Science. The SSRL's work was also supported by the Office Division of Materials Science. The Stanford work was also supported by NSF. A.L. thankd the Instituto Nazionale Fisica della Materia (INFM) and University of Rome "La Sapienza" for support.

**Correspondence and requests for materials should be addressed to: Z-X. Shen (e-mail: zxshen@stanford.edu).**

Figure 1: Ubiquity of a sudden change ("kink") in the dispersion. Top panels are plots of the disepersion (derived from the momentum distribution curves) along (0, 0)- (π, π) (except panel **b** inset, which is off this line) versus the rescaled momentum k' for different samples and at different doping. **a-c**, Doping($\delta$) dependence of LSCO (at 20K; **a**), Bi2212 (superconducting state, 20K; **b**) and Pb-Bi2201 (normal state, 30K; **c**). Dotted lines are guides to the eye. The kink position in **a** is compared with the phonon energy at q=(π, 0) (thick red arrow) and the phonon width and dispersion (shaded area) from neutron data[8]. The doping was determined from the $T_c$ vs doping universal curve. Inset in **b**) dispersions off the (0, 0)- (π, π) direction, showing also a sharpening of the kink on moving away from the nodal direction. The black arrows indicate the position of the kink in the dispersions **d,e**. Temperature dependence of the dispersions for LSCO (**d**, optimally doped) and Bi2221 (**e**, optimally doped). **f**) Doping dependence of λ' (see text) along the (0, 0)- (π, π) direction as a function of doping. Data are shown for LSCO (filled triangles) and NdLSCO (1/8 doping; dilled diamonds), Pb-Bi2201 (filled squares) and Bi2212 (filled circles, in the first Brillouin zone, and unfilled circles in the second zone). The different shadings represent data obtained in different experimental runs. Blue area is a guide to the eyes.

Figure 2: Double peak features in the photoemission spectra due to strong el-ph interaction. Raw energy distribution curves (EDCs) along the (0, 0) to (π, π) direction for the overdoped Pb-Bi2212 (Tc=70K); **a**), for optimally doped Bi2212 (Tc=91K) in panel b) and for underdoped Bi2212 (Tc=84K); **c**). **d**) Raw EDCs for the Be(0001) surface[11]. **e**) EDCs of simulated spectra, obtained in the simple case of an isotropic coupling to a single phonon mode. The straight vertical line highlights the zero energy position and a dashed line highlights the approximate "dip" position, which is related to known phonon energy[8] (see text).

Figure 3: Comparison of the observed effect between photoemission, tunneling and neutron data. **a**) Quasiparticle width along the $\Gamma Y$ ((0, 0)-($\pi$, $\pi$)) (circles) for Bi2212 and Pb-Bi2212, and along $\Gamma M$ ((0, 0)-($\pi$, 0); squares), for the superstructure free Pb-Bi2212 system. **b**) Doping-dependence of the dip position for the Bi2212 samples, near the ($\pi$, 0) region, as indicated in the inset of **b**) by the arrow. The dip position has been obtained both from photoemission (circles) and tunneling data (triangles[28-30]). The photoemission data were collected at SSRL, beam-line 5.4 in transmission mode (I) (green circles) and angular mode (II) (pink circles) and at SSRL, beam line 5.3 (yellow circles). The data at ALS were collected in angular mode (light green circles). Red circles represent the He-lamp data, collected with a Scienta analyzer. Our data are compared with data points from ref. 17 (grey circles). The red arrow indicate the LO phonon energy at ($\pi$, 0)[9], which sets the lower bound for the dip. FWHM, full-width at half maximum; UD, underdoped; OD, overdoped; opt, optimum doping.

Figure 4: Momentum dependence of the velocity ratio above and below the transition temperature. The locations of the sections in the Brillouin zone are shown in the inset of **b**). $\phi$ is defined as the angle between the $\Gamma$-Y line and the line from $\Gamma$ to the point on the Fermi surface considered. **a**), The velocity ratio versus angle $\phi$ along the Fermi surface. The open and filled symbols represent the values obtained fitting the MDCs dispersions in different experimental runs. HT and LT indicate the values of the velocity ratios obtained above and below the transition temperature. The HT data for the Pb-Bi2212 are collected at 100K (red circle) while the LT data at 20K (blue circle). The HT data for the Pb-Bi2201 system (squares) were collected at 30K. The LT data (at 20K) along the nodal direction, for the LSCO and NdLSCO systems are also included (open yellow and filled gray triangles). All velocity ratios are obtained for cuts perpendicular

to the FS at given $\phi$. The inset of **a**) compares the velocity ratios value for the Pb-Bi2212 as obtained from EDC (black filled circles), with data[3] on pure Bi2212 (open diamonds). In **b**) the rescaled MDC-derived dispersions, along the nodal and anti-nodal directions are shown for the Pb-Bi2212. The small deviation above the phonon energy (indicated by the thick arrow), is caused by the difference in dispersion along the two directions, which are quite different if plotted with unscaled k.

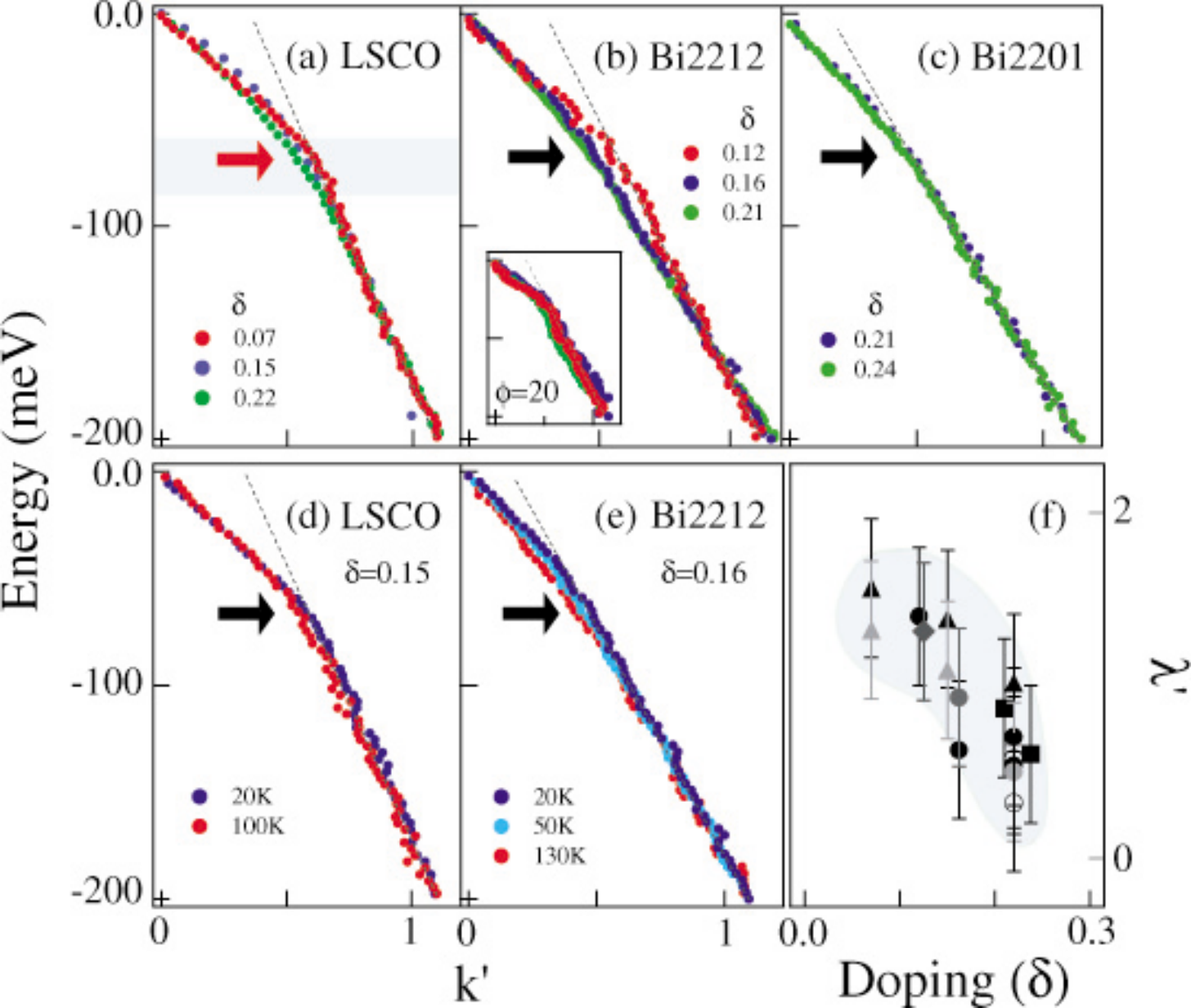

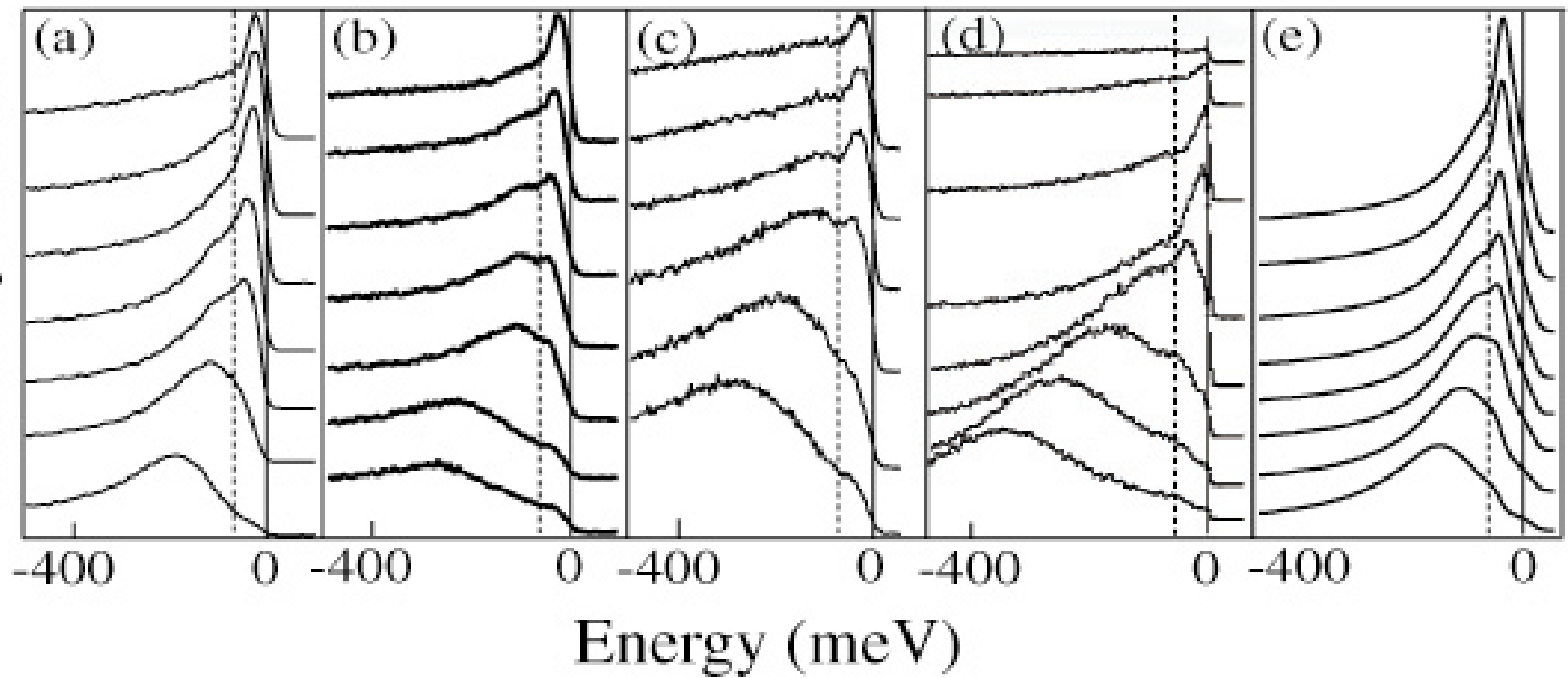

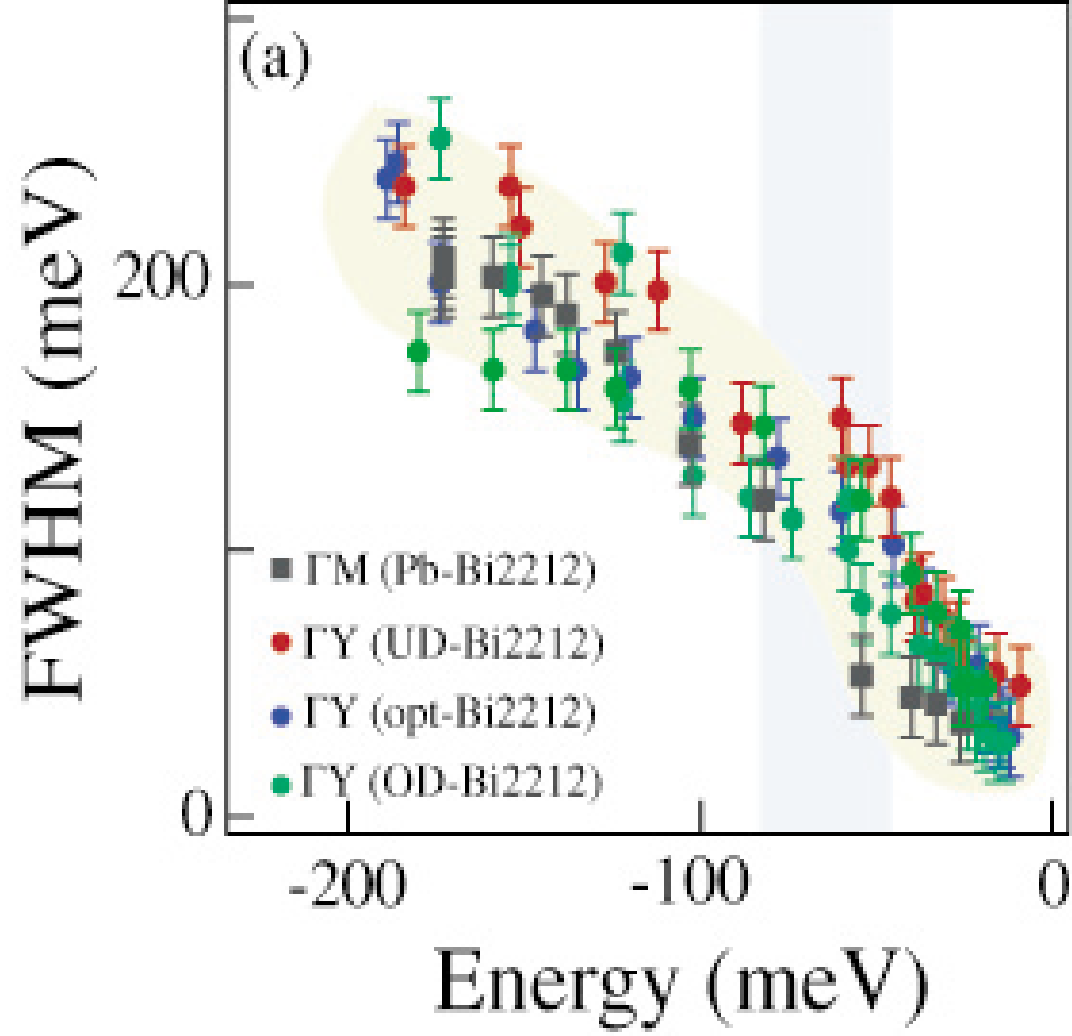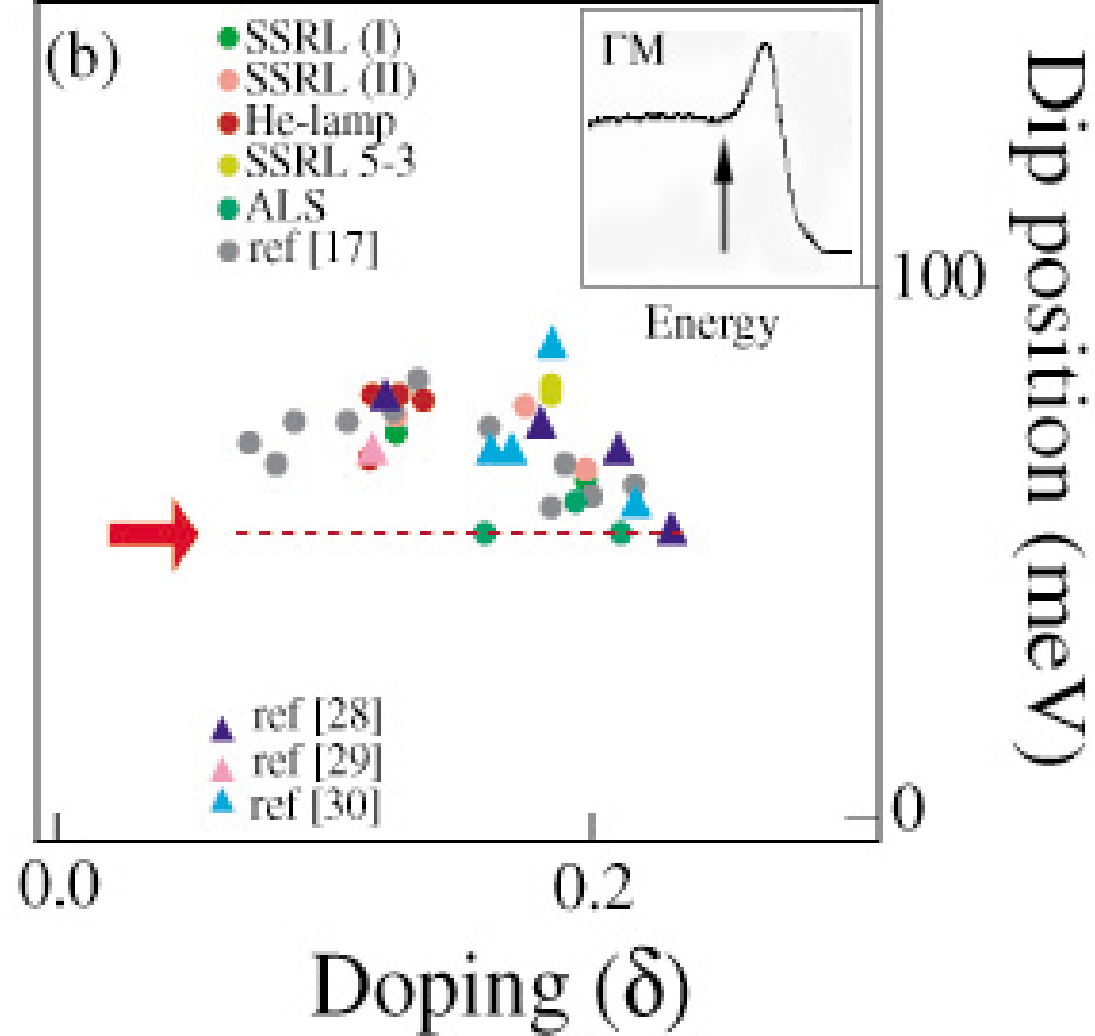

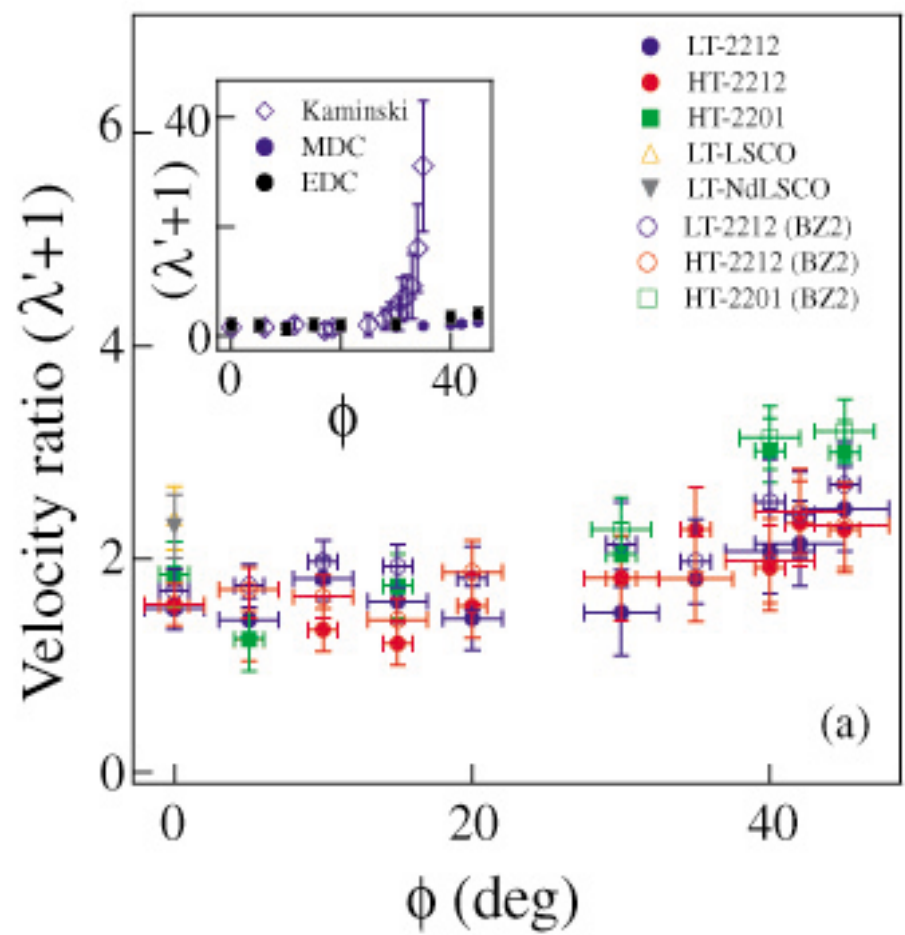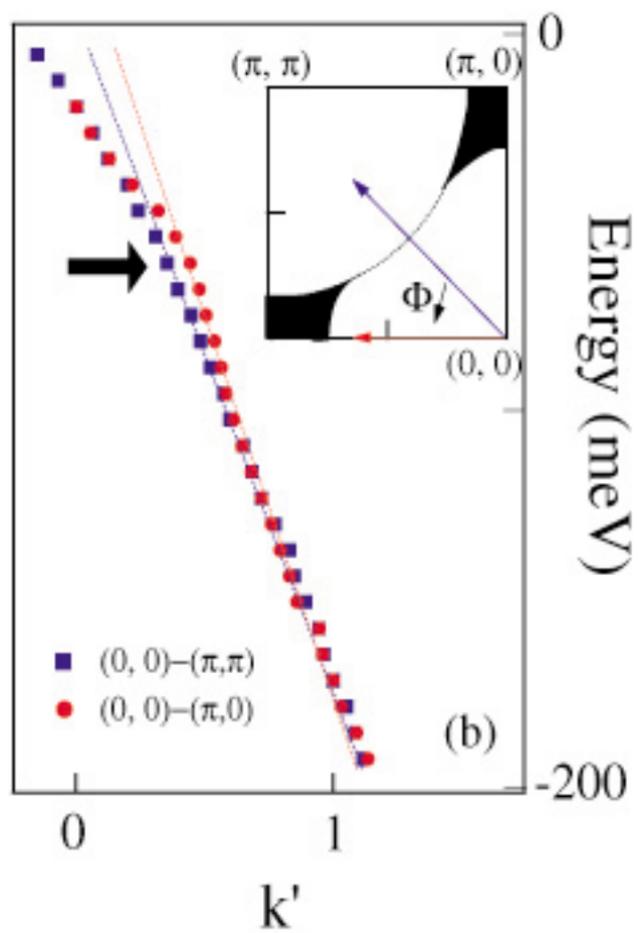